\documentstyle[12pt]{article}
\begin{document}

\def\ds{\displaystyle}
\def\beq{\begin{equation}}
\def\eeq{\end{equation}}
\def\bea{\begin{eqnarray}}
\def\eea{\end{eqnarray}}
\def\beeq{\begin{eqnarray}}
\def\eeeq{\end{eqnarray}}
\def\ve{\vert}
\def\vel{\left|}
\def\ver{\right|}
\def\nnb{\nonumber}
\def\ga{\left(}
\def\dr{\right)}
\def\aga{\left\{}
\def\adr{\right\}}
\def\lla{\left<}
\def\rra{\right>}
\def\rar{\rightarrow}
\def\nnb{\nonumber}
\def\la{\langle}
\def\ra{\rangle}
\def\ba{\begin{array}}
\def\ea{\end{array}}
\def\tr{\mbox{Tr}}
\def\ssp{{\Sigma^{*+}}}
\def\sso{{\Sigma^{*0}}}
\def\ssm{{\Sigma^{*-}}}
\def\xis0{{\Xi^{*0}}}
\def\xism{{\Xi^{*-}}}
\def\qs{\la \bar s s \ra}
\def\qu{\la \bar u u \ra}
\def\qd{\la \bar d d \ra}
\def\qq{\la \bar q q \ra}
\def\gGgG{\la g^2 G^2 \ra}
\def\q{\gamma_5 \not\!q}
\def\x{\gamma_5 \not\!x}
\def\g5{\gamma_5}
\def\sb{S_Q^{cf}}
\def\sd{S_d^{be}}
\def\su{S_u^{ad}}
\def\ss{S_s^{??}}
\def\sbp{{S}_Q^{'cf}}
\def\sdp{{S}_d^{'be}}
\def\sup{{S}_u^{'ad}}
\def\ssp{{S}_s^{'??}}
\def\sig{\sigma_{\mu \nu} \gamma_5 p^\mu q^\nu}
\def\fo{f_0(\frac{s_0}{M^2})}
\def\ffi{f_1(\frac{s_0}{M^2})}
\def\fii{f_2(\frac{s_0}{M^2})}
\def\O{{\cal O}}
\def\sl{{\Sigma^0 \Lambda}}
\def\es{\!\!\! &=& \!\!\!}
\def\ar{&+& \!\!\!}
\def\ek{&-& \!\!\!}
\def\cp{&\times& \!\!\!}
\def\se{\!\!\! &\simeq& \!\!\!}


\renewcommand{\textfraction}{0.2}    
\renewcommand{\topfraction}{0.8}

\renewcommand{\bottomfraction}{0.4}
\renewcommand{\floatpagefraction}{0.8}
\newcommand\mysection{\setcounter{equation}{0}\section}

\def\baeq{\begin{appeq}}     \def\eaeq{\end{appeq}}
\def\baeeq{\begin{appeeq}}   \def\eaeeq{\end{appeeq}}
\newenvironment{appeq}{\beq}{\eeq}
\newenvironment{appeeq}{\beeq}{\eeeq}
\def\bAPP#1#2{
 \markright{APPENDIX #1}
 \addcontentsline{toc}{section}{Appendix #1: #2}
 \medskip
 \medskip
 \begin{center}      {\bf\LARGE Appendix #1 :}{\quad\Large\bf #2}
\end{center}
 \renewcommand{\thesection}{#1.\arabic{section}}
\setcounter{equation}{0}
        \renewcommand{\thehran}{#1.\arabic{hran}}
\renewenvironment{appeq}
  {  \renewcommand{\theequation}{#1.\arabic{equation}}
     \beq
  }{\eeq}
\renewenvironment{appeeq}
  {  \renewcommand{\theequation}{#1.\arabic{equation}}
     \beeq
  }{\eeeq}
\nopagebreak \noindent}

\def\eAPP{\renewcommand{\thehran}{\thesection.\arabic{hran}}}

\renewcommand{\theequation}{\arabic{equation}}
\newcounter{hran}
\renewcommand{\thehran}{\thesection.\arabic{hran}}

\def\bmini{\setcounter{hran}{\value{equation}}
\refstepcounter{hran}\setcounter{equation}{0}
\renewcommand{\theequation}{\thehran\alph{equation}}\begin{eqnarray}}
\def\bminiG#1{\setcounter{hran}{\value{equation}}
\refstepcounter{hran}\setcounter{equation}{-1}
\renewcommand{\theequation}{\thehran\alph{equation}}
\refstepcounter{equation}\label{#1}\begin{eqnarray}}


\newskip\humongous \humongous=0pt plus 1000pt minus 1000pt
\def\caja{\mathsurround=0pt}


\title{
 {\small \begin{flushright}
IPM/P-2007/002\\
\today
\end{flushright}}
       {\Large
                 {\bf
The Effects of Fourth Generation  on the Total Branching Ratio and the Lepton Polarization in
$\Lambda_b \rar \Lambda \ell^+ \ell^-$ decay
                 }
         }
      }

\author{\vspace{1cm}\\
{\small V.Bashiry$^1$\thanks {e-mail: bashiry@ipm.ir} , K.
Azizi$^2$\thanks {e-mail: kazizi@newton.physics.metu.edu.tr}\,\,,
} \\
{\small $^1$ Institute for Studies in Theoretical Physics and
Mathematics (IPM),}\\ {\small P.O. Box 19395-5531, Tehran, Iran
}\\{\small $^2$ Physics Department, Middle East Technical
University, 06531 Ankara, Turkey}}
\date{}
\begin{titlepage}
\maketitle
\thispagestyle{empty}

\begin{abstract}
This study investigates the influence of the fourth generation
quarks on the total branching ratio and the single lepton polarizations in $\Lambda_b \rar \Lambda
\ell^+ \ell^-$ decay. Taking $|V_{t's}V_{t'b}|\sim \{0.01-0.03\}$
with phase just below $90^\circ$, which is consistent with the $b\to
s\ell^+\ell^-$ rate and the $B_s$ mixing parameter $\Delta m_{B_s}$,
we obtain that the total branching ratio and the single lepton($\mu, \, \tau$) polarizations are quite
sensitive to the existence of fourth generation. It can serve as a
good tool to search for new physics effects, precisely, to search
for the fourth generation quarks($t',\, b')$.
\end{abstract}

~~~PACS numbers: 12.60.--i, 13.30.--a, 14.20.Mr
\end{titlepage}

\section{Introduction}
Although the standard model (SM) of electroweak interaction has very
successfully described all existing experimental data, it is
believed that it is a low energy manifestation of some fundamental
theory. Therefore, intensive search for physics beyond the SM is now
being performed in various areas of particle physics. One possible
extension is SM with more than three generations.

 The mass and mixing patterns of the fundamental fermions are the most mysterious aspects
of the particle physics. Even the number of fermion generations is
not fixed by the Standard Model(SM). In this sense, SM may be
treated as an effective theory of fundamental interactions rather
than fundamental particles. The Democratic Mass Matrix approach
\cite{harari}, which is quite natural in the SM framework, may be
considered as the interesting step in true direction. It is
intriguing that Flavors Democracy favors the existence of the fourth
SM family \cite{datta, celikel,Sultansoy:2000dm}. The main
restrictions on the new SM families come from experimental data on
the  $\rho$ and $S$ parameters \cite{Sultansoy:2000dm}. However, the
common mass of the fourth quark ($m_{t'}$) lies between 320 GeV and
730 GeV considering the experimental value of
$\rho=1.0002^{+0.0007}_{-0.0004}$ \cite{PDG}. The last value is
close to upper limit on heavy quark masses, $m_q\leq 700$ GeV
$\approx 4m_t$, which follows from partial-wave unitarity at high
energies \cite{chanowitz}. It should be noted that with preferable
value $a\approx g_w$ Flavor Democracy predicts $m_{t'}\approx 8
m_w\approx 640$ GeV. The above mentioned values for mass of $m_{t'}$
disfavors the fifth SM family both because in general we expect that
$m_t\ll m_{t'}\ll m_{t''}$ and the experimental values of the
$\rho$ and $S$ parameters \cite{Sultansoy:2000dm} restrict the quark
mass up to $700$ GeV.

 Moreover, Democratic Mass Matrix approach provides, in
principle, the possibility to get the small masses
\cite{Hill:1989vn} for the first neutrino species without see-saw
mechanism. The fourth family quarks, if exist, will be copiously
produced at the LHC \cite{Atlas}. Then the fourth family leads to an
essential increase in Higgs boson production cross section via gluon
fusion at hadron colliders \cite{Ginzburg:1998ki}.

 One of the efficient ways to establish the existence of four generation
 is via their indirect manifestations in loop diagrams. Rare decays, induced
  by flavor changing neutral
current (FCNC) $b \rar s(d)$ transitions is at the forefront of our
quest to understand flavor and the origins of CPV, offering one of
the best probes for New Physics (NP) beyond the Standard Model (SM).
Several hints for NP have emerged in the past few years.
For example, a large difference is seen in direct CP asymmetries in
$B\to K\pi$ decays~\cite{HFAG},
\begin{eqnarray}
{\cal A}_{K\pi}
 \equiv A_{\rm CP}(B^0\to K^+\pi^-) = -0.093 \pm 0.015, &&\nonumber\\
{\cal A}_{K\pi^0}
 \equiv A_{\rm CP}(B^+\to K^+\pi^0) = +0.047 \pm 0.026, &&
 \label{data}
\end{eqnarray}
or $\Delta{\cal A}_{K\pi} \equiv {\cal A}_{K\pi^0}-{\cal A}_{K\pi} =
(14\pm 3)\%$~\cite{Barlow}. As this percentage was not predicted
when first measured in 2004, it has stimulated discussion on the
potential mechanisms that it may have been missed in the SM
calculations \cite{BBNS,KLS,BPS05}.

Better known is the mixing-induced CP asymmetry ${\cal S}_f$
measured in a multitude of CP eigenstates $f$. For penguin-dominated
$b \to sq\bar q$ modes, within SM, ${\cal S}_{sq\bar q}$ should be
close to that extracted from $b\to c\bar cs$ modes. The latter is
now measured rather precisely, ${\cal S}_{c\bar cs}=\sin2\phi_1 =
0.674 \pm 0.026$~\cite{Hazumi}, where $\phi_1$ is the weak phase in
$V_{td}$. However, for the past few years, data seem to indicate, at
2.6$\,\sigma$ significance,
\begin{eqnarray}
\Delta {\cal S} \equiv {\cal S}_{sq\bar q}-{\cal S}_{c\bar cs}\leq
0,
 \label{DelS}
\end{eqnarray}
which has stimulated even more discussions.

 Flavor--changing neutral current (FCNC) $b \rar s(d) \ell^+ \ell^-$ decays
provide important tests for the gauge structure of the standard
model (SM) at one--loop level. Moreover, $b \rar s(d) \ell^+ \ell^-$
decay is also very sensitive to the new physics beyond SM. New
physics effects manifest themselves in rare decays in two different
ways, either through new combinations to the new Wilson coefficients
or through the new operator structure in the effective Hamiltonian,
which is absent in the SM. One of the efficient ways in establishing
new physics beyond the SM is the measurement of the lepton
polarization in the inclusive $b\rar s(d)\ell^+
\ell^-$ transition\cite{R4620} and the exclusive $B\rar K(~K^\ast,~\rho,~\gamma)~ \ell^+
\ell^-$ decays \cite{R4621}--\cite{bashirychin}.

In this paper we investigate the possibility of searching for new
physics in the heavy baryon decays $\Lambda_b \rar \Lambda \ell^+
\ell^-$ using the SM with four generations of quarks($b',\, t'$).
The fourth quark ($t'$), like $u,c,t$ quarks, contributes in the $b
\rar s(d) $ transition at loop level. Note that, fourth generation
effects have been widely studied in baryonic and semileptonic B
decays \cite{Hou:2006jy}--\cite{Turan:2005pf}. But, there are few
works related to the exclusive decays $\Lambda_b\rightarrow\Lambda
l^{+}l^{-}$.

The main problem for the description of exclusive decays is to
evaluate the form factors, i.e., matrix elements of the effective
Hamiltonian between initial and final hadron states. It is well
known that in order to describe baryonic   $\Lambda_b \rar \Lambda
\ell^+ \ell^-$ decay a number of form factors are needed (see for
example \cite{R4629}). However, when heavy quark effective theory
(HQET) is applied, only two independent form factors appear
\cite{R46210}.

It should be mentioned here that the exclusive decay $\Lambda_b \rar
\Lambda \ell^+ \ell^-$ decay rate, lepton polarization and heavy($\Lambda_b$)
or light($\Lambda$) baryon polarization(readily measurable) is studied in the SM, the two Higgs
doublet model and using the general form of the effective
Hamiltonian, in \cite{R4629}, \cite{R46211} and
\cite{R46212}--\cite{baryonpol}, respectively.

The sensitivity of the forward--backward asymmetry  to the existence
of fourth generation quarks in the $\Lambda_b \rar \Lambda \ell^+
\ell^-$ decay is investigated in \cite{Turan:2005pf} and it is
obtained that the forward--backward asymmetry is very sensitive to
the fourth generation parameters ($m_{t'}$, $V_{t'b}V^*_{t's}$ ). In
this connection it is natural to ask whether the total branching ratio and the lepton
polarizations are
 sensitive to the fourth generation parameters, in the
"heavy baryon $\rar$ light baryon $\ell^+ \ell^-$" decays. In the
present work we try to answer to this question.

The paper is organized as follows. In section 2, using the effective
Hamiltonian, the general expressions for the longitudinal,
transversal and normal polarizations of leptons are derived. In
section 3 we investigate the sensitivity of these polarizations to
the fourth generation parameters ($m_{t'}$, $V_{t'b}V^*_{t's}$ ).

\section{Lepton polarizations}

The matrix element of the $\Lambda_b \rar \Lambda \ell^+ \ell^-$
decay at quark level is described by $b \rar s \ell^+ \ell^-$
transition for which the effective Hamiltonian at $O(\mu)$ scale can
be written as
 \bea\label{Hgen} {\cal H}_{eff} &=& \frac{4 G_F}{\sqrt{2}}
V_{tb}V_{ts}^\ast \sum_{i=1}^{10} {\cal C}_i(\mu) \, {\cal
O}_i(\mu)~, \eea where the full set of the operators ${\cal
O}_i(\mu)$ and the corresponding expressions for the Wilson
coefficients ${\cal C}_i(\mu)$ in the SM are given in
\cite{R23}--\cite{R24}. As it has already been noted , the fourth
generation is introduced in the same way as three generations in the
SM, and so new operators do not appear and clearly the full operator
set is exactly the same as in SM. The fourth generation changes the
values of the Wilson coefficients $C_7(\mu),~C_9(\mu)$ and
$C_{10}(\mu)$, via virtual exchange of the fourth generation up type
quark $t^\prime$. The above mentioned Wilson coefficients will
modify as \bea\lambda_t C_i \rightarrow \lambda_t
C^{SM}_i+\lambda_{t'} C^{new}_i~,\eea where $\lambda_f=V_{f b}^\ast
V_{fs}$. The unitarity of the $4\times4$ CKM matrix leads to
\bea\lambda_u+\lambda_c+\lambda_t+\lambda_{t'}=0.\eea\ Since
$\lambda_u=V_{ub}^\ast V_{us}$ is very small in strength compared to
the others . Then $\lambda_t\approx -\lambda_c-\lambda_{t'}$ and
$\lambda_c=V_{c b}^\ast V_{cs}\approx 0.04$ is real by convention.
It follows that \bea \lambda_t C^{SM}_i+\lambda_{t'}
C^{new}_i=\lambda_c C^{SM}_i+\lambda_{t'} (C^{new}_i-C^{SM}_i )\eea
It is clear that, for the $m_{t'}\rar m_t$ or $\lambda_{t'}\rar 0$,
$\lambda_{t'} (C^{new}_i-C^{SM}_i )$ term vanishes, as required by
the GIM mechanism. One can also write $C_i$'s in the following form
\bea\label{c4} C_7^{tot}(\mu) &=& C_7^{SM}(\mu) +
\frac{\lambda_{t'}}
{\lambda_t} C_7^{new} (\mu)~, \nnb \\
C_9^{tot}(\mu) &=& C_9^{SM}(\mu) +  \frac{\lambda_{t'}}
{\lambda_t}C_9^{new} (\mu) ~, \nnb \\
C_{10}^{tot}(\mu) &=& C_{10}^{SM}(\mu) +  \frac{\lambda_{t'}}
{\lambda_t} C_{10}^{new} (\mu)~, \eea where the last terms in
these expressions describe the contributions of the $t^\prime$
quark to the Wilson coefficients. $\lambda_{t'}$  can be
parameterized as : \bea {\label{parameter}}
\lambda_{t'}=V_{t^\prime b}^\ast V_{t^\prime
s}=r_{sb}e^{i\phi_{sb}}\eea

 In deriving Eq. (\ref{c4}) we factored
out the term $V_{tb}^\ast V_{ts}$ in the effective Hamiltonian given
in Eq. (\ref{Hgen}). The explicit forms of the $C_i^{new}$ can
easily be obtained from the corresponding expression of the Wilson
coefficients in SM by substituting $m_t \rar m_{t^\prime}$ (see
\cite{R23,R25}). If the $\hat{s}$ quark mass is neglected, the above
effective Hamiltonian leads to following matrix element for the $b
\rar s \ell^+ \ell^-$ decay \bea\label{e1} {\cal H}_{eff} &=&
\frac{G\alpha}{2\sqrt{2} \pi}
 V_{tb}V_{ts}^\ast
\Bigg[ C_9^{tot} \, \bar s \gamma_\mu (1-\gamma_5) b \, \bar \ell
\gamma_\mu \ell + C_{10}^{tot} \bar s \gamma_\mu (1-\gamma_5) b \,
\bar \ell \gamma_\mu \gamma_5 \ell \nnb \\
&-& 2  C_7^{tot}\frac{m_b}{q^2} \bar s \sigma_{\mu\nu} q^\nu
(1+\gamma_5) b \, \bar \ell \gamma_\mu \ell \Bigg]~, \eea where
$q^2=(p_1+p_2)^2$ and $p_1$ and $p_2$ are the final leptons
four--momenta. The effective coefficient $C_9^{tot}$ can be
written in the following form \bea C_9^{tot} = C_9 + Y(s)~, \eea
where $s' = q^2 / m_b^2$ and the function $Y(s')$ contains the
contributions from the one loop matrix element of the four quark
operators. A perturbative calculation leads to the result
\cite{R23,R24}, \bea Y_{per}(s') &=& g(\hat m_c, s') (3 C_1 + C_2
+ 3 C_3 + C_4 + 3 C_5 + C_6) \nnb \\&-& \frac{1}{2} g(1, s') (4
C_3 + 4 C_4 + 3 C_5 + C_6) \nnb \\&-& \frac{1}{2} g(0, s') (C_3 +
3 C_4) + \frac{2}{9} (3 C_3 + C_4 + 3 C_5 + C_6)~, \eea where
$\hat m_c = \frac{m_c}{m_b} $. The explicit expressions for
$g(\hat m_c, s')$, $g(0, s')$, $g(1, s')$ and the values of $C_i$
in the SM can be found in Table 1 \cite{R23,R24}.
\begin{table}
\renewcommand{\arraystretch}{1.5}
\addtolength{\arraycolsep}{3pt}
$$
\begin{array}{|c|c|c|c|c|c|c|c|c|}
\hline C_{1} & C_{2} & C_{3} & C_{4} & C_{5} & C_{6} & C_{7}^{SM} &
C_{9}^{SM} & C_{10}^{SM}\\ \hline
-0.248 & 1.107& 0.011& -0.026& 0.007& -0.031& -0.313& 4.344& -4.669\\
\hline
\end{array}
$$
\caption{The numerical values of the Wilson coefficients at $\mu =
m_{b}$ scale within the SM. The corresponding numerical value of
$C^{0}$ is $0.362$.}
\renewcommand{\arraystretch}{1}
\addtolength{\arraycolsep}{-3pt}
\end{table}

In addition to the short distance contribution, $Y_{per}(s')$
receives also long distance contributions, which have their origin
in the real $c\bar c$ intermediate states, i.e., $J/\psi$,
$\psi^\prime$, $\cdots$. The $J/\psi$ family is introduced by the
Breit--Wigner distribution for the resonances through the
replacement \cite{R26}--\cite{R28} \bea Y(s') = Y_{per}(s') +
\frac{3\pi}{\alpha^2} \, C^{(0)} \sum_{V_i=\psi_i} \kappa_i \,
\frac{m_{V_i} \Gamma(V_i \rar \ell^+ \ell^-)} {m_{V_i}^2 - s' m_b^2
- i m_{V_i} \Gamma_{V_i}}~, \eea where $C^{(0)}= 3 C_1 + C_2 + 3 C_3
+ C_4 + 3 C_5 + C_6$. The phenomenological parameters $\kappa_i$ can
be fixed from ${\cal B} (B \rar K^\ast V_i \rar K^\ast \ell^+
\ell^-) = {\cal B} (B \rar K^\ast V_i)\, {\cal B} ( V_i \rar \ell^+
\ell^-)$, where the data for the right hand side is given in
\cite{R29}. For the lowest resonances $J/\psi$ and $\psi^\prime$ one can
use $\kappa = 1.65$ and $\kappa = 2.36$, respectively (see \cite{R30}).

The amplitude of the exclusive $\Lambda_b \rar \Lambda\ell^+ \ell^-$
decay can be obtained by sandwiching ${\cal H}_{eff}$ for the $b
\rar s \ell^+ \ell^-$ transition between initial and final baryon
states, i.e., $\lla \Lambda \vel {\cal H}_{eff} \ver \Lambda_b
\rra$. It follows from Eq. (\ref{e1}) that in order to calculate the
$\Lambda_b \rar \Lambda\ell^+ \ell^-$ decay amplitude the following
matrix elements are needed \bea &&\lla \Lambda \vel \bar s
\gamma_\mu (1 \mp \gamma_5) b \ver \Lambda_b
\rra~,\nnb \\
&&\lla \Lambda \vel \bar s \sigma_{\mu\nu} (1 \mp \gamma_5) b \ver
\Lambda_b
\rra~,\nnb \\
&&\lla \Lambda \vel \bar s (1 \mp \gamma_5) b \ver \Lambda_b
\rra~.\nnb \eea Explicit forms of these matrix elements in terms of
the form factors are presented in \cite{R46212} (see also
\cite{R4629}). The matrix element of the $\Lambda_b \rar
\Lambda\ell^+ \ell^-$ can be written as \bea \label{e3} \lefteqn{
{\cal M} = \frac{G \alpha}{4 \sqrt{2}\pi} V_{tb}V_{ts}^\ast \Bigg\{
\bar \ell \gamma^\mu \ell \, \bar u_\Lambda \Big[ A_1 \gamma_\mu
(1+\gamma_5) +
B_1 \gamma_\mu (1-\gamma_5) }\nnb \\
\ar i \sigma_{\mu\nu} q^\nu \big[ A_2 (1+\gamma_5) + B_2
(1-\gamma_5) \big] +q_\mu \big[ A_3 (1+\gamma_5) + B_3 (1-\gamma_5)
\big]\Big] u_{\Lambda_b}
\nnb \\
\ar \bar \ell \gamma^\mu \gamma_5 \ell \, \bar u_\Lambda \Big[E_1
\gamma_\mu (1-\gamma_5)  + i\sigma_{\mu\nu} q^\nu E_2 (1-\gamma_5) +
E_3 q^{\mu}(1-\gamma_5)\big] \Big] u_{\Lambda_b}\Bigg\}~, \eea where
$P=p_{\Lambda_b}+ p_\Lambda$. Explicit expressions of the functions
$A_i,~B_i,$ and $E_i$ $(i=1,2,3)$  are given as follows
\cite{R46212}: \bea A_1&=&-\frac{4 m_b}{m_{\Lambda_b}}~F_2~
C^{tot}_7\nnb\\A_2&=&-\frac{4 m_b}{q^2}~(F_1+\sqrt{r}F_2)~
C^{tot}_7\nnb\\A_3&=&-\frac{4 m_b
m_{\Lambda}}{q^2m_{\Lambda_b}}~F_2~
C^{tot}_7\nnb\\B_1&=&2(F_1+\sqrt{r}F_2)~C^{tot}_9\nnb\\B_2&=&\frac{2
F_2}{m_{\Lambda_b}}~C^{tot}_9\nnb\\B_3&=&\frac{4 m_b}{q^2}~F_2~
C^{tot}_7\nnb\\E_1&=&2(F_1+\sqrt{r}F_2)~
C^{tot}_{10}\nnb\\E_2&=&E_3=\frac{2F_2}{m_{\Lambda_b}}~C^{tot}_{10}\eea

From the expressions of the above-mentioned matrix elements Eq.
(\ref{e3}) we observe that $\Lambda_b \rar\Lambda \ell^+\ell^-$
decay is described in terms of many form factors. When HQET is
applied to the number of independent form factors, as it has already
been noted,  reduces to two ($F_1$ and $F_2$) irrelevant with the
Dirac structure of the corresponding operators and it is obtained in
\cite{R46210} that \bea \label{e4} \lla \Lambda(p_\Lambda) \vel \bar
s \Gamma b \ver \Lambda(p_{\Lambda_b}) \rra = \bar u_\Lambda
\Big[F_1(q^2) + \not\!v F_2(q^2)\Big] \Gamma u_{\Lambda_b}~, \eea
where $\Gamma$ is an arbitrary Dirac structure,
$v^\mu=p_{\Lambda_b}^\mu/m_{\Lambda_b}$ is the four--velocity of
$\Lambda_b$, and $q=p_{\Lambda_b}-p_\Lambda$ is the momentum
transfer. Comparing the general form of the form factors with
(\ref{e5}), one can easily obtain the following relations among them
(see also \cite{R4629}) \bea \label{e5}
g_1 \es f_1 = f_2^T= g_2^T = F_1 + \sqrt{r} F_2~, \nnb \\
g_2 \es f_2 = g_3 = f_3 = g_T^V = f_T^V = \frac{F_2}{m_{\Lambda_b}}~,\nnb \\
g_T^S \es f_T^S = 0 ~,\nnb \\
g_1^T \es f_1^T = \frac{F_2}{m_{\Lambda_b}} q^2~,\nnb \\
g_3^T \es \frac{F_2}{m_{\Lambda_b}} \ga m_{\Lambda_b} + m_\Lambda \dr~,\nnb \\
f_3^T \es - \frac{F_2}{m_{\Lambda_b}} \ga m_{\Lambda_b} - m_\Lambda
\dr~, \eea where $r=m_\Lambda^2/m_{\Lambda_b}^2$.

Having obtained the matrix element for the $\Lambda_b \rar\Lambda
\ell^+ \ell^-$ decay, we next aim to calculate lepton polarizations
with the help of this matrix element. We write the $\ell^\mp$ spin
four--vector in terms of a unit vector $\vec{\xi}_\mp$ along the
$\ell^\mp$ momentum in its rest frame as \bea \label{e6} s_\mu^\mp =
\ga \frac{\vec{p}^{\;\mp}\cdot \vec{\xi}^{\;\mp}}{m_\ell},
\vec{\xi}^{\;\mp} + \frac{\vec{p}^{\;\mp} (\vec{p}^{\;\mp} \cdot
\vec{\xi}^{\;\mp})}{E_\ell+m_\ell} \dr ~, \eea and choose the unit
vectors along the longitudinal, normal and transversal components of
the $\ell^-$ polarization to be \bea \label{e7} \vec{e}_L^{\;\mp} =
\frac{\vec{p}^{\;\mp}}{\vel \vec{p}^- \ver}~, ~~~ \vec{e}_N^{\;\mp}
= \frac{\vec{p}_\Lambda\times \vec{p}^{\;\mp}} {\vel
\vec{p}_\Lambda\times \vec{p}^- \ver}~,~~~ \vec{e}_T^{\;\mp} =
\vec{e}_N^{\;\mp} \times \vec{e}_L^{\mp}~, \eea respectively, where
$\vec{p}^{\;\mp}$ and $\vec{p}_{\Lambda}$ are the three momenta of
$\ell^\mp$ and $\Lambda$, in the center of mass frame of the $\ell^+
\ell^-$ system. Obviously, $\vec{p}^{\;+}=-\vec{p}^{\;-}$ in this
reference frame.

The differential decay rate of the $\Lambda_b \rar \Lambda \ell^+
\ell^-$ decay for any spin direction $\vec{\xi}^{\;\mp}$ can be
written as \bea \label{e8} \frac{d\Gamma(\vec{\xi}^{\;\mp})}{ds} =
\frac{1}{2} \ga \frac{d\Gamma}{ds}\dr_0 \Bigg[ 1 + \Bigg( P_L^\mp
\vec{e}_L^{\;\mp} + P_N^\mp \vec{e}_N^{\;\mp} + P_T^\mp
\vec{e}_T^{\;\mp} \Bigg) \cdot \vec{\xi}^{\;\mp} \Bigg]~, \eea where
$\ga d\Gamma/ds \dr_0$ corresponds to the unpolarized differential
decay rate, $s=q^2/m_{\Lambda_b}^2$ and $P_L^\mp$, $P_N^\mp$ and
$P_T^\mp$ represent the longitudinal, normal and transversal
polarizations of $\ell^\mp$, respectively. The unpolarized decay
width  in Eq. (\ref{e8}) can be written as

\bea \label{e9} \ga \frac{d \Gamma}{ds}\dr_0 = \frac{G^2
\alpha^2}{8192 \pi^5} \vel V_{tb} V_{ts}^\ast \ver^2
\lambda^{1/2}(1,r,s) v \Big[{\cal T}_0(s) +\frac{1}{3} {\cal T}_2(s)
\Big]~, \eea where $\lambda(1,r,s) = 1 + r^2 + s^2 - 2 r - 2 s - 2
rs$ is the triangle function and $v=\sqrt{1-4m_\ell^2/q^2}$ is the
lepton velocity. The explicit expressions for ${\cal T}_0$ and
${\cal T}_2$ are given by: \bea {\cal T}_0&=&4m_{\Lambda_b}^2\Big\{
8m_{\ell}^2 m_{\Lambda_b}^2\hat{s}(1+r-\hat{s})|E_3|^2+16m_{\ell}^2
m_{\Lambda_b}\sqrt{r}(1-r+\hat{s})\mbox{\rm Re}[E_1^\ast
E_3]+\nnb\\&&8(2m_{\ell}^2+
m_{\Lambda_b}^2\hat{s})\{(1-r+\hat{s})m_{\Lambda_b}\sqrt{r}\mbox{\rm
Re}[A_1^\ast A_2+B_1^\ast B_2]
-\nnb\\&&m_{\Lambda_b}(1-r-\hat{s})\mbox{\rm Re}[A_1^\ast
B_2+A_2^\ast B_1]-2\sqrt{r}(\mbox{\rm Re}[A_1^\ast
B_1]+m_{\Lambda_b}^2\hat{s}\mbox{\rm Re}[A_2^\ast B_2])
\}+\nnb\\&&2\Big(
4m_{\ell}^2(1+r-\hat{s})+m_{\Lambda_b}^2\Big[(1-r)^2-\hat{s}^2\Big]\Big)\Big(|A_1|^2+|B_1|^2\Big)+\nnb\\&&
2m_{\Lambda_b}^2\Big(4m_{\ell}^2\Big[\lambda+(1+r-\hat{s})\hat{s}\Big]+
m_{\Lambda_b}^2\hat{s}\Big[(1-r)^2-\hat{s}^2\Big]\Big)\Big(|A_2|^2+|B_2|^2
\Big)-\nnb\\&&2\Big(4m_{\ell}^2(1+r-\hat{s})-m_{\Lambda_b}^2\Big[(1-r)^2-\hat{s}^2\Big]\Big)|E_1|^2+\nnb\\&&
2 m_{\Lambda_b}^3\hat{s}v^2\Big(4(1-r+\hat{s})\sqrt{r}\mbox{\rm
Re}[E_1^\ast
E_2]-m_{\Lambda_b}\Big[(1-r)^2-\hat{s}^2\Big]|E_2|^2\Big)\Big\} \eea
\bea {\cal
T}_2=-8m_{\Lambda_b}^4v^2\lambda\Big(|A_1|^2+|B_1|^2+|E_1|^2-m_{\Lambda_b}^2\hat{s}(|A_2|^2+|B_2|^2+|E_2|^2)\Big).\eea
The polarizations $P_L$, $P_N$ and $P_T$ are defined as: \bea
P_i^{(\mp)}(q^2) = \frac{\ds{\frac{d \Gamma}{ds}
                   (\vec{\xi}^{\;\mp}=\vec{e}_i^{\;\mp}) -
                   \frac{d \Gamma}{ds}
                   (\vec{\xi}^{\;\mp}=-\vec{e}_i^{\;\mp})}}
              {\ds{\frac{d \Gamma}{ds}
                   (\vec{\xi}^{\;\mp}=\vec{e}_i^{\;\mp}) +
                  \frac{d \Gamma}{ds}
                  (\vec{\xi}^{\;\mp}=-\vec{e}_i^{\;\mp})}}~, \nnb
\eea where $i = L,N,T$. $P_L$ and $P_T$ are $P$--odd, $T$--even,
while $P_N$ is $P$--even, $T$--odd and $CP$--odd. The explicit forms
of the expressions for the longitudinal $P_L$, transversal $P_T$ and
normal $P_N$ lepton polarizations are as follows:
 \bea\label{pL} P^{\mp}_L&=&\Big\{
-32 m_{\ell}~m_{\Lambda_b}^3
v(1-\sqrt{r})[\hat{s}-(1+\sqrt{r})^2]\mbox{\rm Re}[E_1^\ast
F_1]\pm128m_{\Lambda_b}^4 \hat{s}v\sqrt{r}\mbox{\rm Re}[A_1^\ast
E_1]\nnb\\&& \mp64m_{\Lambda_b}^5\sqrt{r}(1-r+\hat{s})\hat{s}v
\mbox{\rm Re}[B_1^\ast E_2+B_2^\ast
E_1]\pm128m_{\Lambda_b}^6\hat{s}^2v \sqrt{r}\mbox{\rm Re}[A_2^\ast
E_2]\nnb\\&&-16m_{\Lambda_b}^4\hat{s}v[\hat{s}-(1+\sqrt{r})^2]
\mbox{\rm Re}[F_1^\ast F_2+2m_{\ell}~E_3^*F_1]\nnb\\&&
\pm64m_{\Lambda_b}^5\hat{s}v (1-r-\hat{s})\mbox{\rm Re}[A_1^\ast
E_2+A_2^\ast E_1]\nnb\\&&\mp\frac{64}{3}m_{\Lambda_b}^4v
[1+r^2+r(\hat{s}-2)+\hat{s}(1-2\hat{s})]\mbox{\rm Re}[B_1^\ast E_1]
\nnb\\&&\mp\frac{64}{3}m_{\Lambda_b}^6v\hat{s}
[2+r(2r-4-\hat{s})-\hat{s}(1+\hat{s})]\mbox{\rm Re}[B_2^\ast
E_2]\Big\}\Big{/}\Big({\cal
T}_0(\hat{s})+\frac{1}{3}{\cal{T}}_2(\hat{ s})\Big),\eea
 \bea \label{pT} P^{\mp}_T&=&\Big\{ -16 \pi
m_{\ell}~m_{\Lambda_b}^3\sqrt{\hat{s}\lambda}(|A_1|^2-|B_1|^2)+
32\pi m_{\ell}~m_{\Lambda_b}^4\sqrt{\hat{s}\lambda}~\mbox{\rm
Re}[A_1^\ast B_2-A_2^\ast B_1]\nnb\\&& \mp16 \pi
m_{\ell}~m_{\Lambda_b}^4\sqrt{\hat{s}\lambda}~\mbox{\rm Re}[A_1^\ast
E_3-A_2^\ast E_1]\mp4 \pi
m_{\Lambda_b}^4\sqrt{\hat{s}\lambda}(1+\sqrt{r})~\mbox{\rm
Re}[(A_1+B_1)^\ast F_2]\nnb\\&& +16 \pi
m_{\ell}~m_{\Lambda_b}^5\sqrt{\hat{s}\lambda}(1-r)(|A_2|^2-|B_2|^2)\mp16\pi
m_{\ell}~m_{\Lambda_b}^3\sqrt{\frac{\lambda}{\hat{s}}}(1-r)\mbox{\rm
Re}[B_1^\ast E_1]\nnb\\&&+ 16\pi
m_{\ell}~m_{\Lambda_b}^4\sqrt{r\hat{s}\lambda}~\mbox{\rm
Re}[2A_1^\ast A_2-2B_1^\ast B_2\mp B_1^\ast E_3\mp B_2^\ast
E_1]\nnb\\&& \pm 4 \pi
m_{\Lambda_b}^5\hat{s}\sqrt{\hat{s}\lambda}\Big\{\mbox{\rm
Re}[(A_2+B_2)^\ast F_2]+4m_{\ell}~\mbox{\rm Re}[B_2^\ast
E_3]\Big\}\\&& +4\pi
m_{\Lambda_b}^5\hat{s}\sqrt{\lambda\hat{s}}~v^2\mbox{\rm
Re}[E_2^\ast F_1]-4\pi
m_{\Lambda_b}^4\sqrt{\lambda\hat{s}}~v^2(1+\sqrt{r})\mbox{\rm
Re}[E_1^\ast F_1]\Big\}\Big{/}\Big({\cal
T}_0(\hat{s})+\frac{1}{3}{\cal{T}}_2(\hat{ s})\Big),\nnb\eea

\bea \label{pN} P^{\mp}_N&=&\Big\{\mp 16 \pi
m_{\ell}~m_{\Lambda_b}^3 v\sqrt{\hat{s}\lambda}~\mbox{\rm
Im}[B_1^\ast E_1]\pm 16 \pi m_{\ell}~m_{\Lambda_b}^4
v\sqrt{\hat{s}\lambda}~\mbox{\rm Im}[A_2^\ast E_1-A_1^\ast
E_2]\nnb\\&&-4 \pi~m_{\Lambda_b}^4
v\sqrt{\hat{s}\lambda}(1+\sqrt{r})~\mbox{\rm Im}[\pm(A_1+B_1)^\ast
F_1+ E_1^\ast F_2]+4 m_{\ell}~\mbox{\rm Im}[E_2^\ast E_3]\nnb\\&&
\pm16\pi m_{\ell}~m_{\Lambda_b}^5
v\sqrt{\hat{s}\lambda}(1-r)~\mbox{\rm Im}[B_2^\ast E_2]+4 \pi
~m_{\Lambda_b}^5 v\hat{s}\sqrt{\hat{s}\lambda}~\mbox{\rm
Im}[\pm(A_2+B_2)^\ast F_1+ E_2^\ast F_2]\nnb\\&&-16 \pi
m_{\ell}~m_{\Lambda_b}^4 v\sqrt{r\hat{s}\lambda}~\mbox{\rm
Im}[E_2^\ast(\pm ~B_1+E_1)+ E_1^\ast(\pm
~B_2+E_3)]\Big\}\Big{/}\Big({\cal
T}_0(\hat{s})+\frac{1}{3}{\cal{T}}_2(\hat{s})\Big).\eea
 where the --(+) sign in
these formulas corresponds to the particle (antiparticle),
respectively.

 It follows from Eq. (\ref{pL}) that the
difference between $P_L^-$ and $P_L^+$, in massless lepton case, is
the same as SM with three generations because it depends on form
factors $F_1$ and $F_2$ . Again in the same way, in massless lepton
case, the difference between $P_N^-$ and $P_N^+$ depends on  fourth
generation CP violation phase($\phi_{sb}$) and $r_{sb}$. We get the
result of SM with three generations if $\phi_{sb}$ is zero.

Combined analysis of the lepton and antilepton polarizations can
give additional information about the existence of new physics,
since in the SM, $P_L^-+P_L^+=0,~P_N^-+P_N^+=0$ and
$P_T^--P_T^+\simeq 0$ (in $m_\ell \rar 0$ limit). Therefore, if
nonzero values for the above mentioned combined asymmetries are
measured in the experiments, it can be considered as an unambiguous
indication of the existence of new physics. But looking at Eq.
(\ref{pL}), we see that $P_L^-+P_L^+$ is the same as SM result in
$m_\ell \rar 0$ limit. Therefore, in order to look for fourth
generation effects we can look at $~P_N^-+P_N^+$ and $~P_T^--P_T^+$
in $m_\ell \rar 0$ limit. The nonzero values of above mentioned
quantities will indicate the existence of fourth generation effects.

\section{Numerical analysis}

In this section we will study the dependence of the total
branching ratio and lepton polarizations as well as combined
lepton polarization to the fourth quark mass($m_{t'}$) and the
product of quark mixing matrix elements ($V_{t^\prime b}^\ast
V_{t^\prime s}=r_{sb}e^{i\phi_{sb}}$). The main input parameters
in the calculations are the form factors. Since the literature
lacks exact calculations for the form factors of the $\Lambda_b
\rar \Lambda$ transition, we will use the results from QCD sum
rules approach in combination with HQET \cite{R46210, R46214},
which reduces the number of quite many form factors into two. The
$\hat{s}$ dependence of these form factors can be represented in
the following way \bea F(q^2) = \frac{F(0)}{\ds 1-a_F s + b_F
s^2}~, \nnb \eea where parameters $F_i(0),~a$ and $b$ are listed
in Table 2.
\begin{table}
\renewcommand{\arraystretch}{1.5}
\addtolength{\arraycolsep}{3pt}
$$
\begin{array}{|l|ccc|}
\hline & F(0) & a_F & b_F \\ \hline F_1 &
\phantom{-}0.462 & -0.0182 & -0.000176 \\
F_2 & -0.077 & -0.0685 &\phantom{-}0.00146 \\ \hline
\end{array}
$$
\caption{Transition form factors for $\Lambda_b \rar \Lambda \ell^+
\ell^-$ decay in the QCD sum rules method.}
\renewcommand{\arraystretch}{1}
\addtolength{\arraycolsep}{-3pt}
\end{table}

We use the next--to--leading order logarithmic approximation for the
resulting values of the Wilson coefficients $C_9^{eff},~C_7$ and
$C_{10}$ in the SM \cite{R46215,R46216} at the re--normalization
point $\mu=m_b$. It should be noted that, in addition to short
distance contribution, $C_9^{eff}$ receives also long distance
contributions from the real $\bar c c$ resonant states of the
$J/\psi$ family. In the present work we do not take  the long
distance effects into account. The input parameters we used in this
analysis are as follows:\\
  $ m_{\Lambda_b}=5.624$GeV, $m_{\Lambda}=1.116$GeV, $m_b=4.8$GeV,
  $m_c=1.35$GeV, $m_{\tau}=1.778$GeV,\\ $m_{\mu}=0.105$GeV,
$\lambda_c=0.045$, $\alpha^{-1}=129$, $G_F=1.166\times
10^{-5}$GeV$^{-2}$\\
 In order to perform quantitative analysis of the
total branching ratio and the lepton polarizations the values of
the new parameters($m_{t'},\,r_{sb},\,\phi_{sb}$) are needed.
Using the experimental values of $B\rar X_s \gamma$ and $B\rar X_s
\ell^+ \ell^-$, the bound on $r_{sb}\sim\{0.01-0.03\}$ has been
obtained \cite{Arhrib:2002md} for $\phi_{sb}\sim\{0-2\pi\}$ and
$m_{t'}\sim\{300,400\}~$(GeV). We are doing complete analysis
about the range of the new parameters considering the recent
experimental value of the  ${\cal{B}}_r (B\rar X_s \ell^+
\ell^-=(1.59\pm0.5)\times 10^{-6})$\cite{HFAG}. Right now, we have
obtained that in the case of the $1\sigma$ level deviation from
the measured branching ratio the maximum values of $m_{t'}$ are
below than the theoretical upper limits The results shown in Table
3 \cite{FZVB}.
\begin{table}
\renewcommand{\arraystretch}{1.5}
\addtolength{\arraycolsep}{3pt}
$$
\begin{array}{|c|c|c|c |c|}
\hline  r_{sb} & 0.005 & 0.01 &0.02 & 0.03 \\
\hline
m_{t'}(GeV) &511 &373 & 289 & 253\\
\hline
\end{array}
$$
\caption{The experimental limit on $ m_{t'} $ for
$\phi_{sb}=\pi/2$}
\renewcommand{\arraystretch}{1}
\addtolength{\arraycolsep}{-3pt}
\end{table}
 In the
foregoing numerical analysis we vary $m_{t'}$ in the range $175\le
m_{t'} \le 600$GeV. The lower range is because of the fact that the
fourth generation up quark should be heavier than the third
ones($m_t \leq m_{t'}$)\cite{Sultansoy:2000dm}. The upper range comes
from the experimental bounds on the $\rho$ and $S$ parameters of SM,
which we mentioned above(see Introduction). We took $r_{sb}\sim
\{0.01-0.03\}$ with phase around $90^\circ$($\phi_{sb}\approx
90^\circ$), which are consistent with the $b\to s\ell^+\ell^-$ rate
and the $B_s$ mixing parameter $\Delta
m_{B_s}$\cite{Hou:2006jy,glenzinski}.

Before performing numerical analysis, few words about lepton
polarizations are in order. From explicit expressions of the lepton
polarizations one can easily see that they depend on both $\hat{s}$ and
the new parameters($m_{t'},\,r_{sb}$). We should eliminate the
dependence of the lepton polarization on one of the variables. We
eliminate the variable $\hat{s}$ by performing integration over $\hat{s}$ in the
allowed kinematical region. The total branching ratio and the
averaged lepton polarizations are defined as \bea {\cal B}_r&=&
\int_{4 m_\ell^2/m_{\Lambda_b}^2}^{(1-\sqrt{r})^2}
 \frac{d{\cal B}}{d\hat{s}} d\hat{s},
\nnb\\\lla P_i \rra &=& \frac{ \int_{4
m_\ell^2/m_{\Lambda_b}^2}^{(1-\sqrt{r})^2} P_i \frac{d{\cal B}}{d\hat{s}}
d\hat{s}} {{\cal{B}}_r}~. \eea

The dependence of the total branching ratio and lepton polarizations
$\lla P_L^-\rra, ~\lla P_T^-\rra$, $\lla P_N^-\rra$, $\lla
P_T^--P_T^+\rra$ and $\lla P_N^-+P_N^+\rra$ on the new
parameters($m_{t'},\,r_{sb}$) are shown in Figs (1)--(9). From these
figures we obtain the following results.

\begin{itemize}

\item ${\cal B}_r$  strongly depends on the fourth quark mass($m_{t'}$)
 and the product of quark mixing matrix elements($r_{sb}$) for both $\mu$
 and $\tau$ channels. Furthermore, for both channels, ${\cal B}_r$ is an
 increasing function of both $m_{t'}$ and $r_{sb}$.

\item Although, $\lla P_L^-\rra$ and $\lla
P_T^-\rra$ strongly  depends on the fourth quark mass($m_{t'}$) and
the product of quark mixing matrix elements($r_{sb}$) for both $\mu$
and $\tau$ channels. But, its magnitude is a decreasing function of
both $m_{t'}$ and $r_{sb}$. So, the existence of fourth generation of
quarks will suppress the magnitude of $\lla P_L^-\rra$ and $\lla
P_T^-\rra$.

\item The normal polarization following from Eq. (\ref{pN}) is proportional
to the imaginary parts of the combination of the products of the
Wilson coefficients, $m_{t'}$ and $r_{sb}$. There are two different
 contributions to the non--zero value of $\lla P_N^-\rra$. First, is
 due to the imaginary part of the $C_9^{eff}$, while the second, is due
to $\phi_{sb}$, which we assume to be $\approx
90^\circ$ in this work, and which
 therefore makes a purely imaginary contribution to Eq. (\ref{c4}).
Moreover, since $\lla P_N^-\rra$ is proportional to the
lepton mass, for the $\mu$ channel it is negligible  in the SM3
 and the SM4($\lla P_N^-\rra_{max}\sim 1\%$). For the $\tau$ case,
 where $\lla P_N^-\rra\sim 1\%$ in the SM, it shows stronger
dependence on $(m_{t'},~r_{sb})$. It is interesting to note that $\lla
P_N^-\rra\sim 5\%$  at $300\leq m_{t'}\leq 400$(GeV). Therefore, measurement of the $\lla
P_N^-\rra$ for $\tau$ channel can serve as good clue for existence
of fourth generation of quarks.

\end{itemize}

Our numerical analysis for the combined lepton and antilepton
polarizations leads to the following results.

\begin{itemize}

\item In the $m_\ell \rar 0$ limit, $\lla P_L^- + P_L^+ \rra$
practically coincides with SM result.

For $\tau$ case, $\lla P_L^- + P_L^+ \rra$ exhibits strong
dependence only on $m_{\tau}$, rather than the new
parameters($m_{t'},~r{sb})$ and practically there is no difference
between the result of SM with three generations.

\item Situation for the combined $\lla P_T^- - P_T^+ \rra$
polarization is as follows.\\
For the $\mu$ channel, $\lla P_T^- - P_T^+ \rra$ is approximately
zero in the SM3 and the SM4.

In the $\tau$ case, $\lla P_T^- - P_T^+ \rra$ is observed to be
strongly dependent on new parameters ($m_{t'}, r_{sb}$). $\lla P_T^- -
P_T^+ \rra$ is decreasing for increasing values of both ($m_{t'},
r_{sb}$). The magnitude of $\lla P_T^- - P_T^+ \rra$ for the $\tau$
channel lies in the region $(0.25 \div 0.85)$ depending on the
variations of the ($m_{t'}, r_{sb}$).

\item Numerical calculations show that the combined $\lla P_N^- +
P_N^+ \rra$ polarization, exhibits strong dependence
 on the ($m_{t'},~r_{sb}$).

 $\lla P_N^- +P_N^+ \rra$ is approximately zero for $\mu$ channels in the
SM, but considering the SM with four generations, it receives the
maximum value of around $1\%$ at $300\leq m_{t'}\leq 400$(GeV). It may be hard for it to be measured in future
experiments i.e., at LHC, unless a large
amount of $\Lambda_b$ (i.e., $\sim 10^{12}$) are created. But, measurement of non zero value($\sim 1\%$) of $\lla
P_N^- +P_N^+ \rra$ for $\mu$ case will be the direct indication of
new physics effects.

The situation for $\tau$ case is more interesting. The considerable
($\sim 6$ times) enhancement can be seen at $300\leq m_{t'}\leq
400$(GeV) in the magnitude of $\lla P_N^- +P_N^+ \rra$ . The
measurement of $\lla P_N^- +P_N^+ \rra$ for $\tau $ case can serve
as a good tool when looking for the fourth generation of quarks.

\end{itemize}

From these analyzes we can conclude that the measurement of the
magnitude of not only the total branching ratio but also $\lla P_i^-
\rra$ and $\lla P_i^- +(-) P_i^+\rra$ (-- sign is for the
transversal polarization case) is an indication of the existence of
new physics beyond the SM.

In conclusion, we present the analysis of the total
branching ratio and the lepton polarizations in the exclusive
$\Lambda_b \rar \Lambda \ell^- \ell^+$ decay, by using the SM with
four generations of quarks. The sensitivity of the total branching ratio, longitudinal,
transversal and normal polarizations of $\ell^-$, as well as
lepton--antilepton combined asymmetries on the new parameters that
come out of fourth generations, are studied. We find out that both
the total branching ratio and the lepton polarizations show a strong
dependence on the fourth quark ($m_{t'}$) and the product of quark
mixing matrix elements ($V_{t^\prime b}^\ast V_{t^\prime
s}=r_{sb}e^{i\phi_{sb}}$). The results can serve as a good tool to
look for physics beyond the SM.

\section{Acknowledgment}
The authors would like to thank T. M. Aliev for his useful
discussions.

\newpage

\section*{Figure captions}

{\bf Fig. (1)} The dependence of the branching ratio for the
$\Lambda_b \rar \Lambda \mu^+ \mu^-$ decay on the fourth generation
quark mass $m_{t'}$ for three different values of
 $r_{sb}$.\\ \\
{\bf Fig. (2)} The same as in Fig. (1), but for the $\tau$ lepton.\\ \\
{\bf Fig. (3)} The dependence of the longitudinal lepton
polarization asymmetry for the $\Lambda_b \rar \Lambda \mu^+ \mu^-$
decay on the fourth generation quark mass $m_{t'}$ for three different
values of $r_{sb}$.\\ \\
{\bf Fig. (4)}
The same as in Fig. (3), but for the $\tau$ lepton.\\
\\
 {\bf Fig. (5)} The dependence of the transversal lepton
polarization asymmetry for the $\Lambda_b \rar \Lambda \mu^+ \mu^-$
decay on the fourth generation quark mass $m_{t'}$ for three different
values of $r_{sb}$.\\ \\
{\bf Fig. (6)} The same as in Fig. (5), but for the $\tau$ lepton.\\
\\
 {\bf Fig. (7)}The dependence of the normal lepton
polarization asymmetry for the $\Lambda_b \rar \Lambda \tau^+ \tau^-$
decay on the fourth generation quark mass $m_{t'}$ for three different
values of $r_{sb}$.\\ \\
{\bf Fig. (8)} The dependence of the combined normal lepton
polarization asymmetry for the $\Lambda_b \rar \Lambda \tau^+ \tau^-$
decay on the fourth generation quark mass $m_{t'}$ for three different
values of $r_{sb}$.\\ \\
{\bf Fig. (9)} The dependence of the combined transversal lepton
polarization asymmetry for the $\Lambda_b \rar \Lambda \tau^+
\tau^-$ decay on the fourth generation quark mass $m_{t'}$ for three
different values of $r_{sb}$.

\newpage

\begin{figure}
\vskip 1.5 cm
    \includegraphics{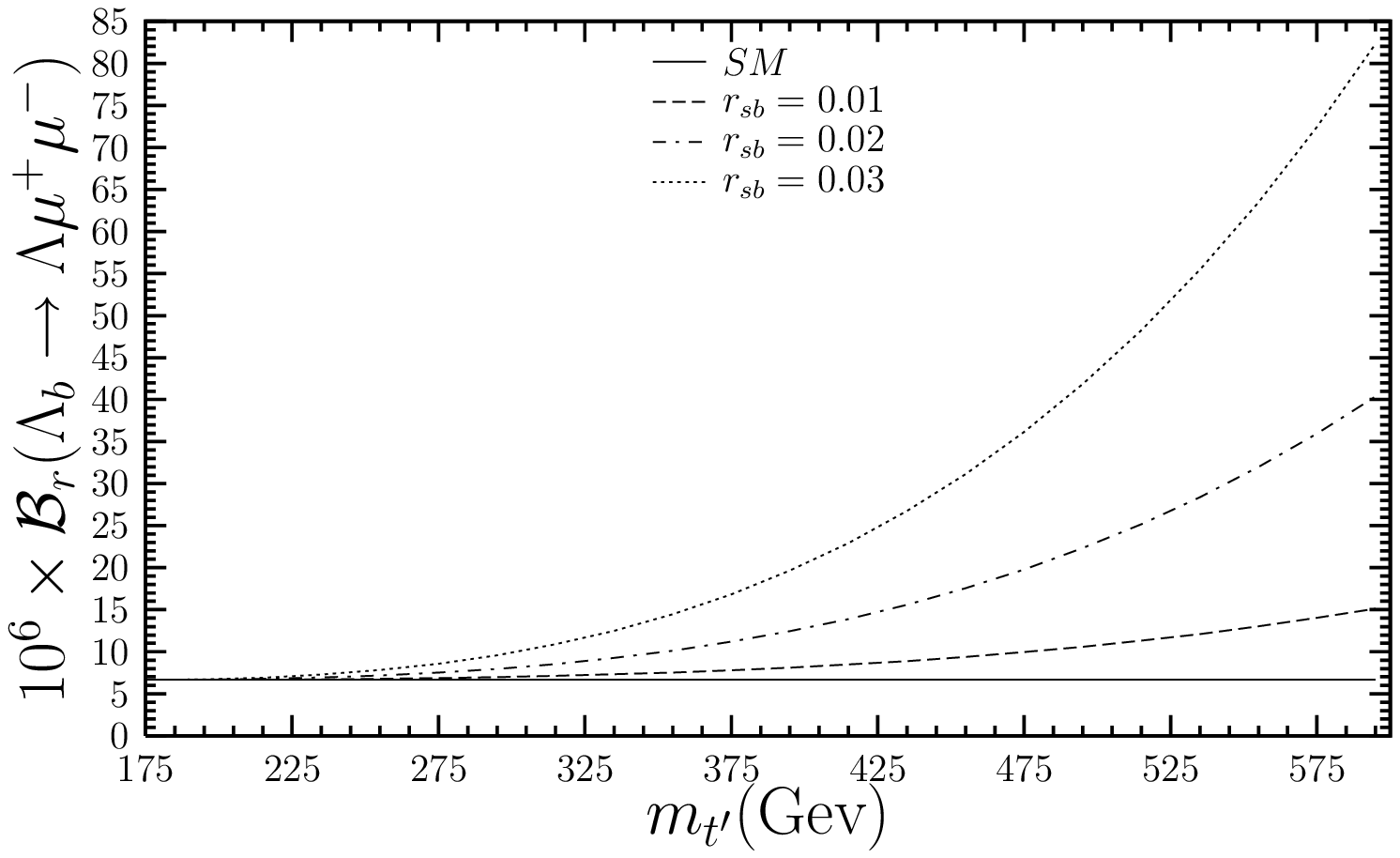}
\vskip 7.8cm
\caption{}
\end{figure}

\begin{figure}
\vskip 2.5 cm
    \includegraphics{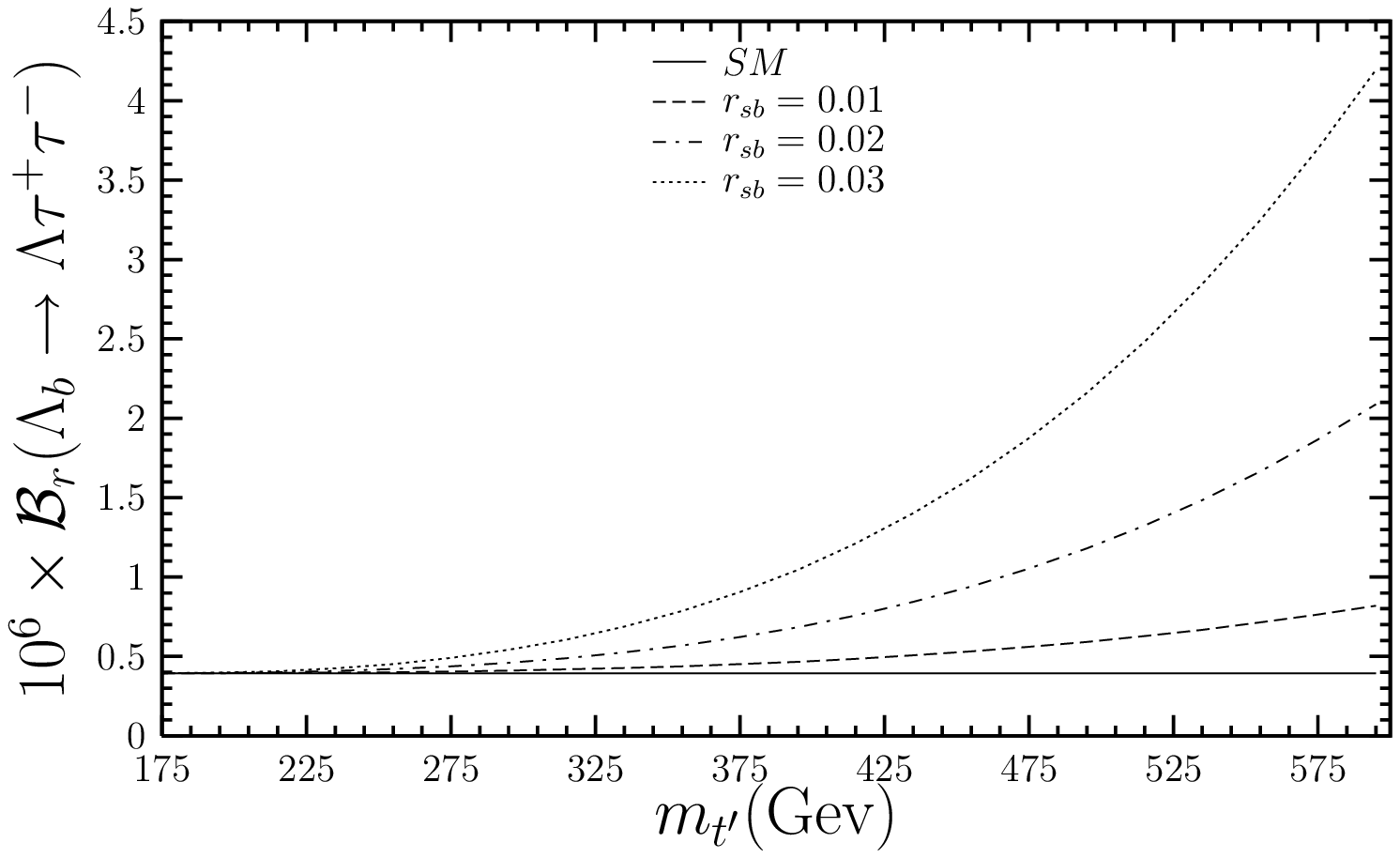}
\vskip 7.8 cm
\caption{}
\end{figure}

\begin{figure}
\vskip 1.5 cm
    \includegraphics{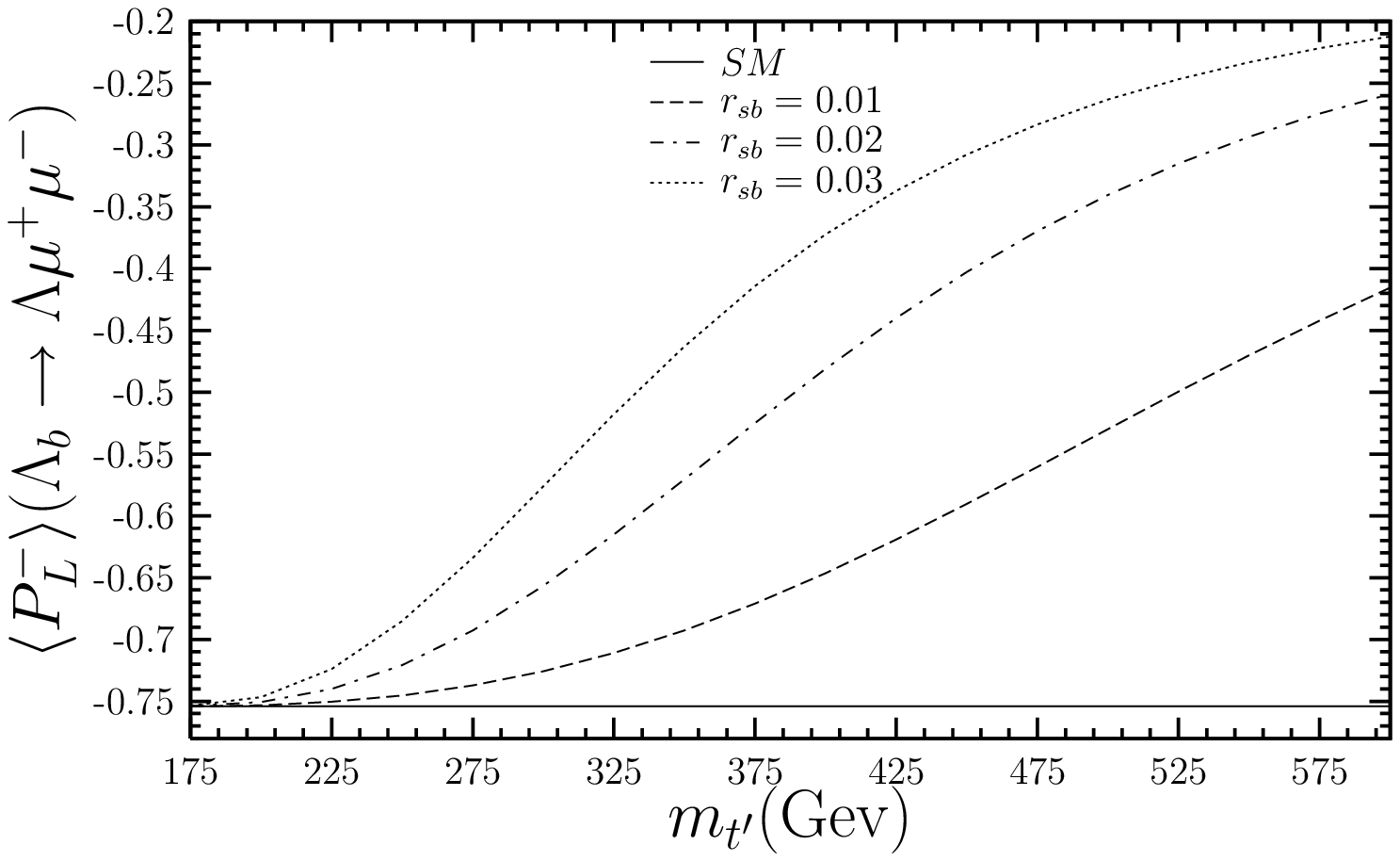}
\vskip 7.8cm
\caption{}
\end{figure}

\begin{figure}
\vskip 2.5 cm
    \includegraphics{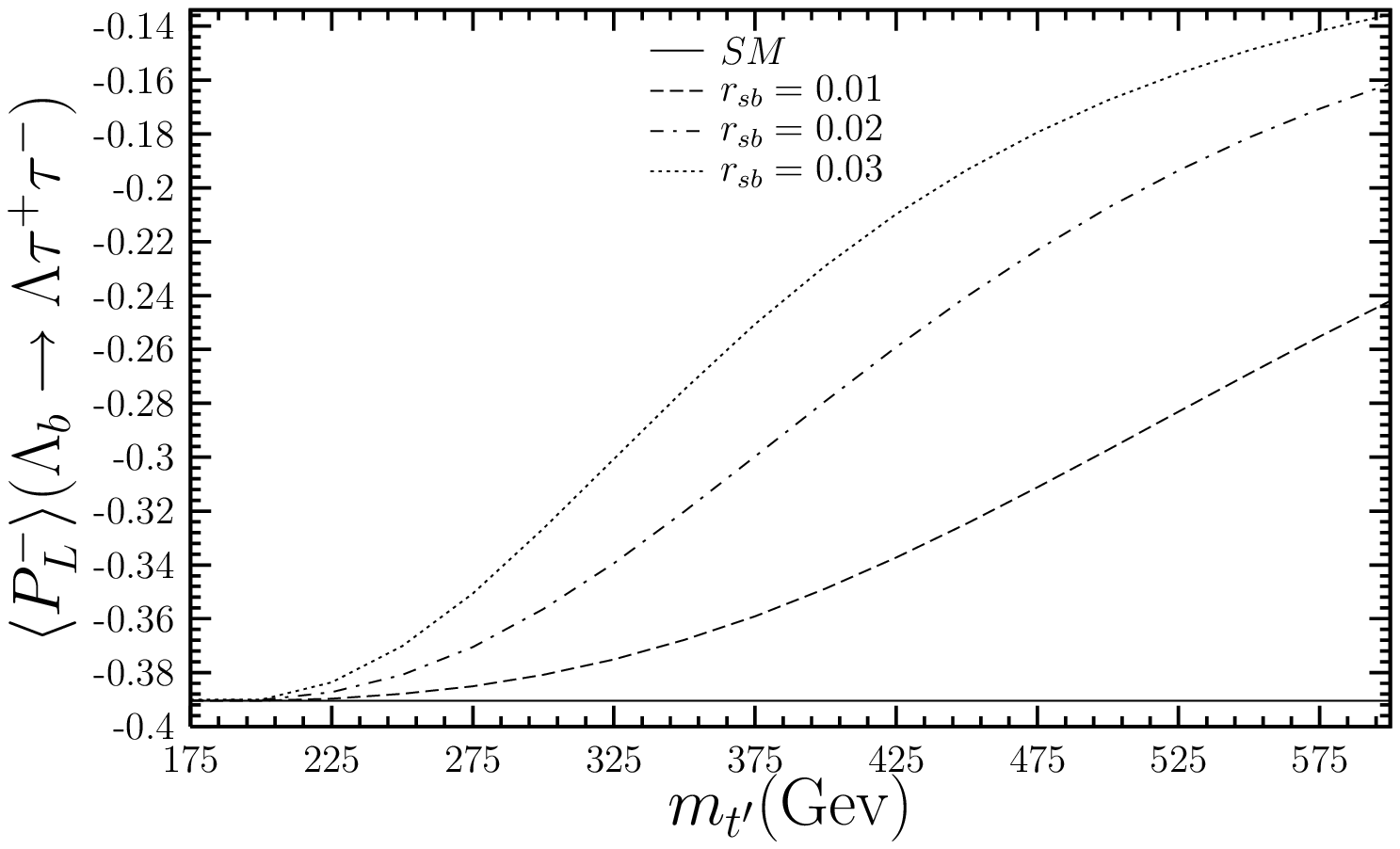}
\vskip 7.8 cm
\caption{}
\end{figure}

\begin{figure}
\vskip 1.5 cm
    \includegraphics{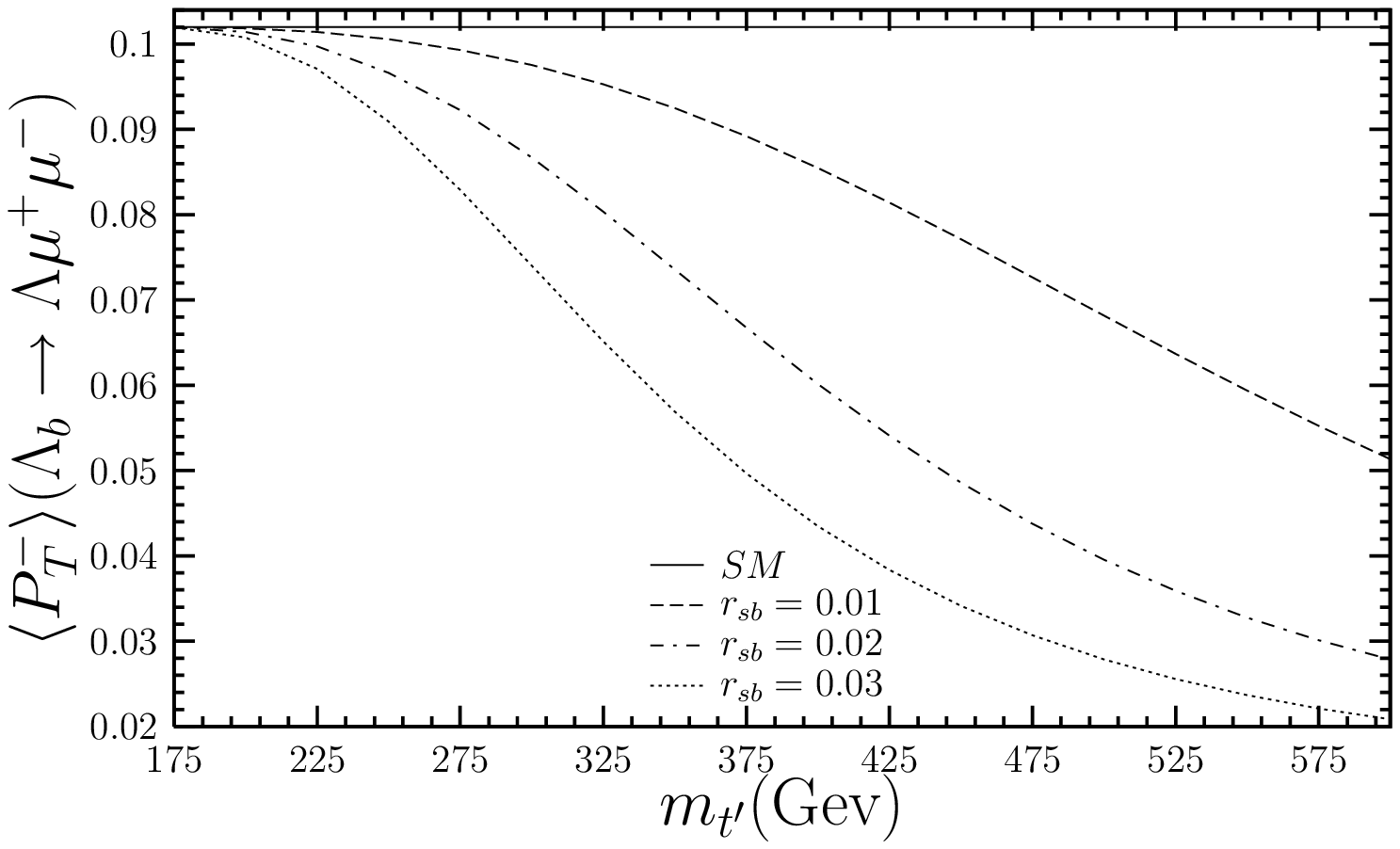}
\vskip 7.8cm
\caption{}
\end{figure}

\begin{figure}
\vskip 2.5 cm
    \includegraphics{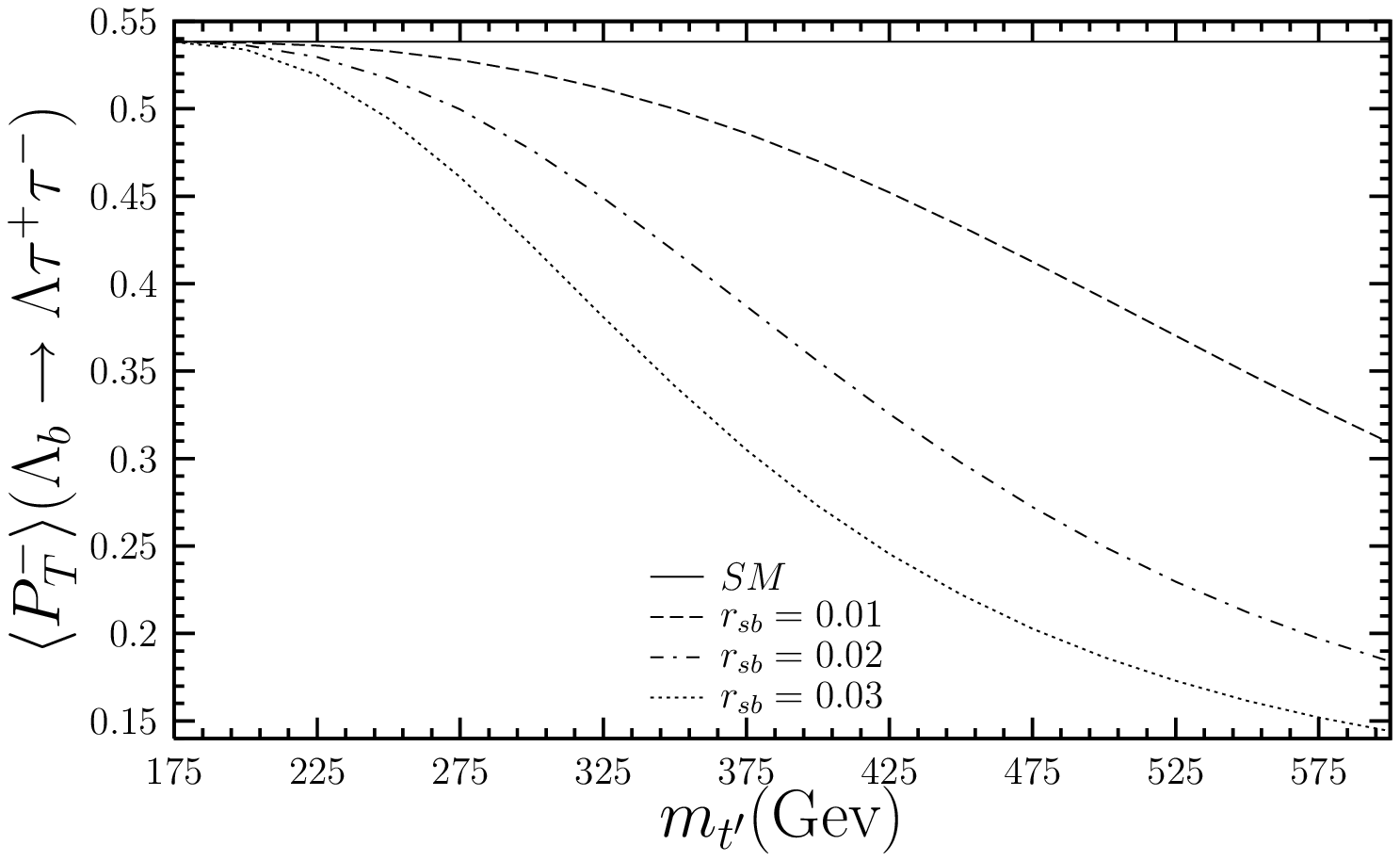}
\vskip 7.8 cm
\caption{}
\end{figure}

\begin{figure}
\vskip 2.5 cm
    \includegraphics{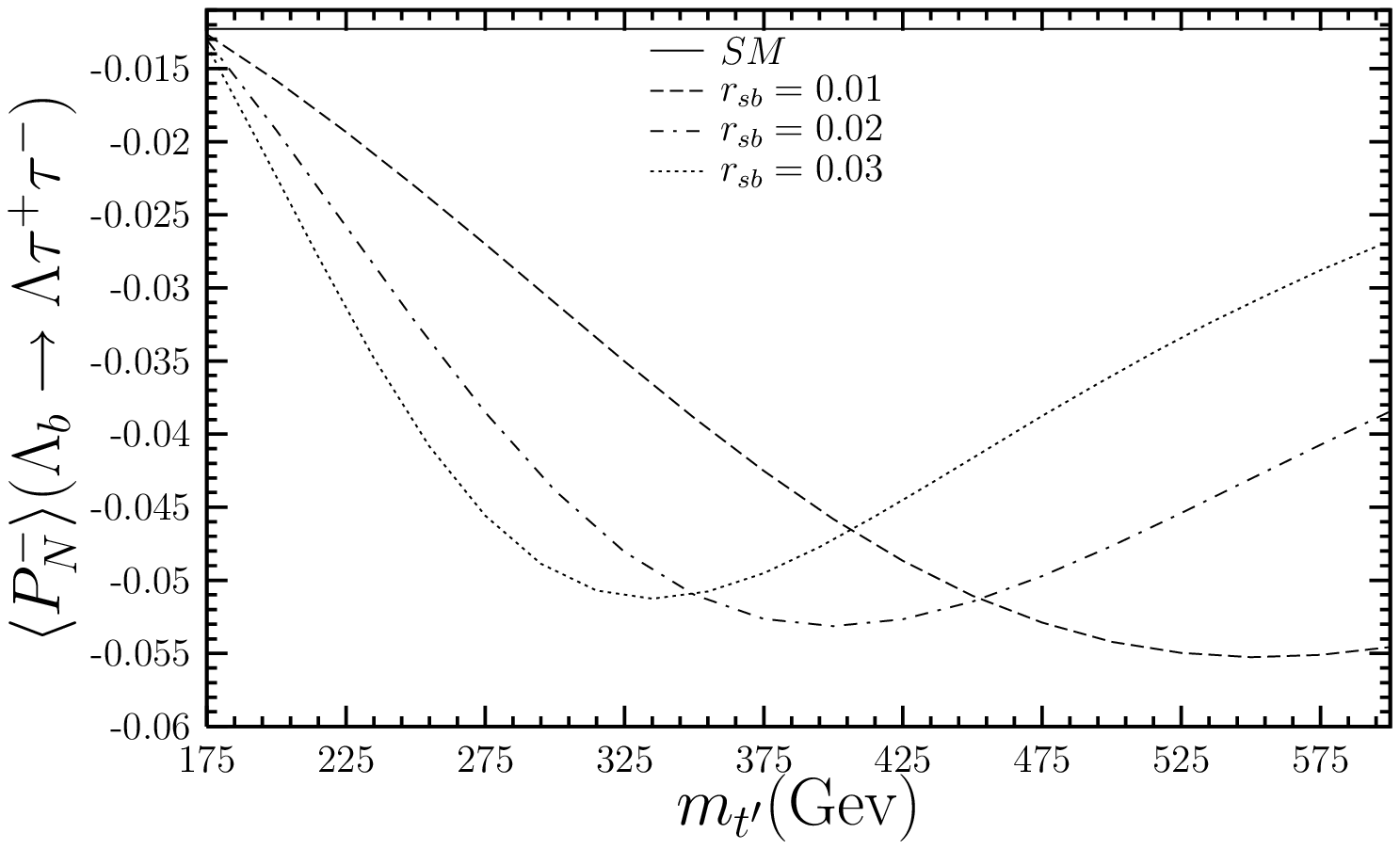}
\vskip 7.8 cm
\caption{}
\end{figure}

\begin{figure}
\vskip 2.5 cm
    \includegraphics{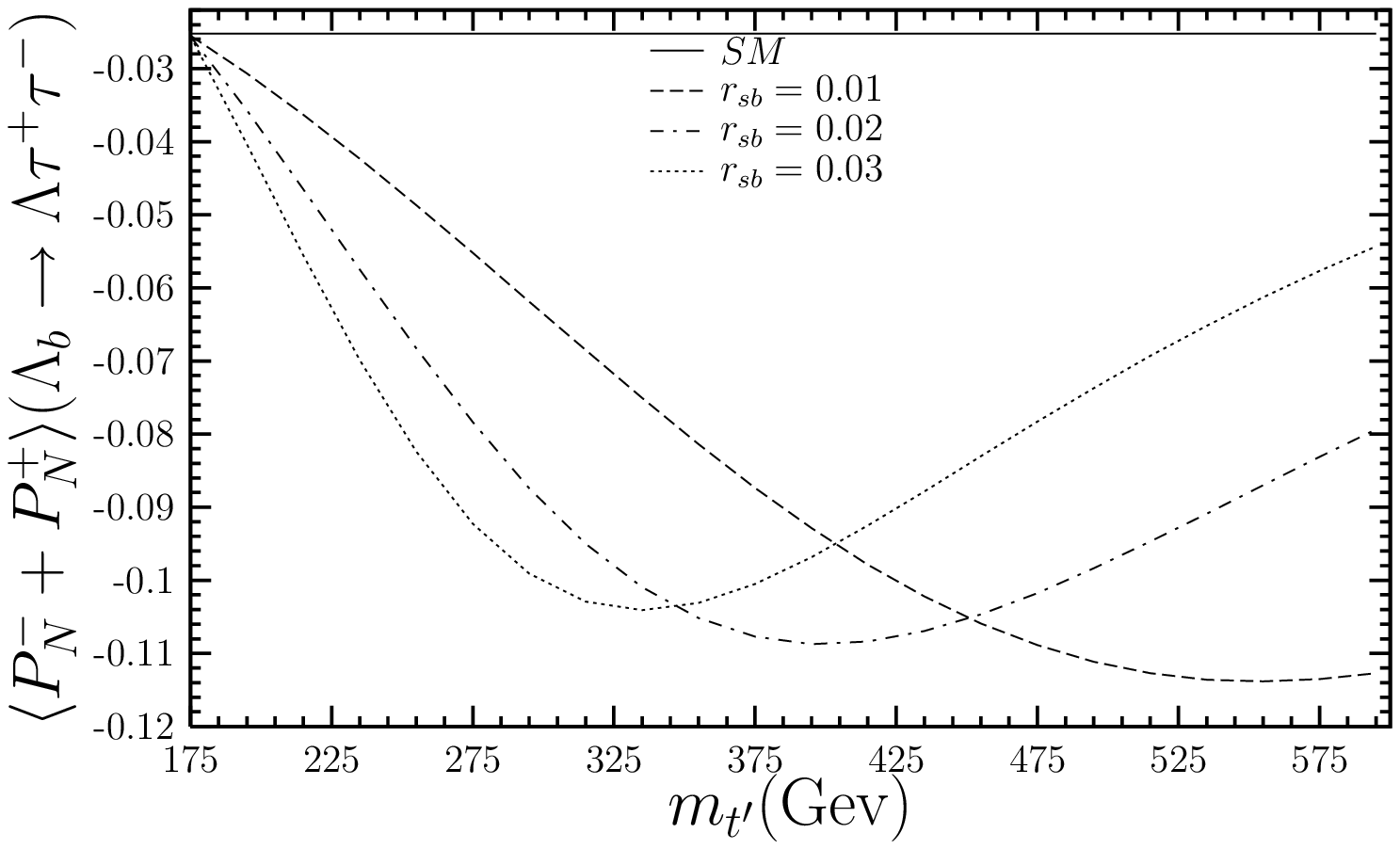}
\vskip 7.8 cm
\caption{}
\end{figure}

\begin{figure}
\vskip 1.5 cm
    \includegraphics{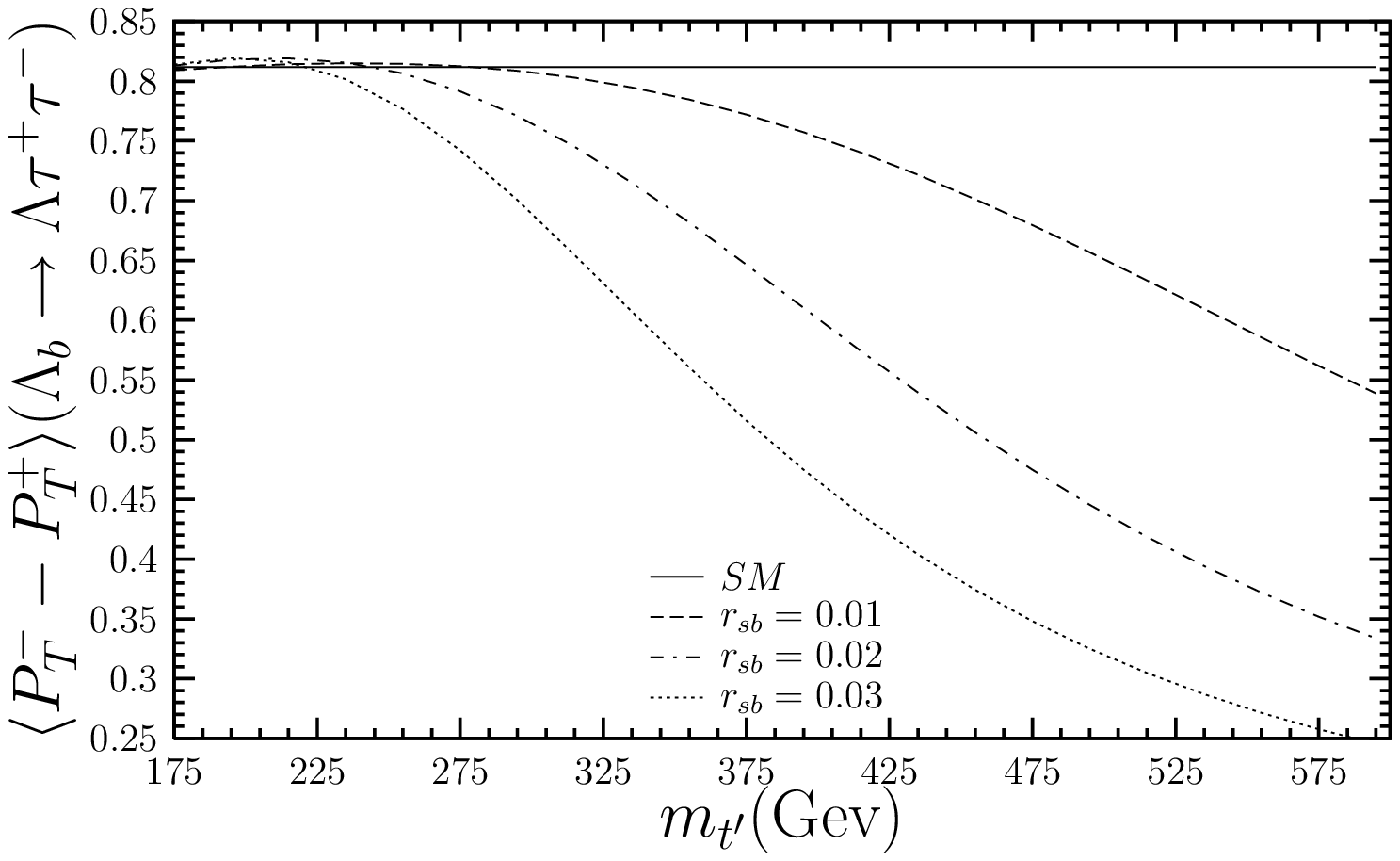}
\vskip 7.8cm
\caption{}
\end{figure}

\end{document}